\documentclass[journal]{IEEEtran}
\ifCLASSINFOpdf
\else
\fi

\usepackage{amssymb}
\usepackage{amsthm}
\usepackage{epsfig}
\usepackage{cite}

\begin{document}
%
\title{Protecting Against Untrusted Relays: An Information Self-encrypted Approach}
%
%
%

\author{Hao Niu, Yao Sun,
       \ and Kaoru~Sezaki
\thanks{H. Niu, Y. Sun and K. Sezaki are with the University of Tokyo.}
}
\maketitle

\begin{abstract}
The reliability and transmission distance are generally limited for the wireless communications due to the severe channel fading. As an effective way to resist the channel fading, cooperative relaying is usually adopted in wireless networks where neighbouring nodes act as relays to help the transmission between the source and the destination. Most research works simply regard these cooperative nodes trustworthy, which may be not practical in some cases especially when transmitting confidential information. In this paper, we consider the issue of untrusted relays in cooperative communications and propose an information self-encrypted approach to protect against these relays. Specifically, the original packets of the information are used to encrypt each other as the secret keys such that the information cannot be recovered before all of the encrypted packets have been received. The information is intercepted only when the relays obtain all of these encrypted packets. It is proved that the intercept probability is reduced to zero exponentially with the number of the original packets. However, the security performance is still not satisfactory for a large number of relays. Therefore, the combination of destination-based jamming is further adopted to confuse the relays, which makes the security performance acceptable even for a large number of relays. Finally, the simulation results are provided to confirm the theoretical analysis and the superiority of the proposed scheme.
\end{abstract}


%
\IEEEpeerreviewmaketitle

\section{Introduction}

The widespread use of wireless devices motivates more research attention on the technologies of wireless communications. Compared to the cable systems, wireless communications suffers a lot from the channel fading. Diversity technique is thus usually used to resist the channel fading, e.g., bit-interleaving in time domain and multiple-input and multiple-output (MIMO) in spatial domain. From the pioneering work of \cite{Sendonaris,Laneman}, the cooperative communications has been studied extensively as an emerging scheme to harvest the spatial diversity gain, which does not need multiple antennas at the devices and thus is more flexible. By exploiting other nodes in the network to act as the relays, multiple transmission links can be provided to improve the transmission reliability. In addition, multi-hop transmission can be also performed through the cooperation to realize the communication between the source and the destination, when they are not in the coverage area of each other.

Exploiting cooperative communications to improve the transmission security is also a heating topic in terms of physical layer security (PLS). Different from the cryptograph, PLS utilizes the channel fading to realize perfect secrecy from the perspective of information theory \cite{Wyner,Csiszar,Barros,Gopala,Saad,Yang}. Briefly speaking, the confidential information can be transmitted securely at a secrecy rate, which is limited by the difference between the channel capacities of the legitimate link and the eavesdropping link. The maximum achievable secrecy rate is called as secrecy capacity. The PLS realizes perfect secrecy even if the eavesdropper has infinite computing power. The research works \cite{Krikidis1,Dong,Bassily,Hui,Krikidis2,Zou1,Fan,Zou2} analyze how to enhance PLS, i.e., improve the secrecy capacity or reduce the secrecy outage probability (the probability that the instantaneous secrecy capacity is less than a target secrecy rate), through the cooperative communications. Among them, \cite{Krikidis1,Dong,Bassily,Hui} combine relay selection with cooperative jamming while \cite{Krikidis2,Zou1,Fan,Zou2} only employ relay selection technique.

However, all of these works assume that the cooperative relays are trustworthy and do not consider the information leakage at the relays. In practical networks, the relays may be untrusted. That is, they are willing to provide their resources to assist the transmission but at the same time they also intend to decode the transmitted information. In \cite{He1} and \cite{He2}, the authors study this issue and propose a destination-based jamming (DBJ) scheme to achieve a positive secrecy capacity in the cooperative networks with an untrusted relay. \cite{Jeong} considers the MIMO nodes and designs a joint secure beamforming scheme to resist the untrusted relay. In \cite{Huang}, the secrecy outage probabilities with an untrusted relay is analyzed for different cooperation schemes. The optimal power allocation is employed in \cite{Wang} for the secure transmission in untrusted relay systems. In addition, references \cite{Sun1,Khodakarami,Sun2} further analyzes the cooperative communications with multiple untrusted relays.

The works in \cite{He1,He2,Jeong,Huang,Wang,Sun1,Khodakarami,Sun2} focus on the perfect secrecy through the PLS, which is not necessary in many applications. Therefore, we exploit the fountain codes for secure wireless transmission in \cite{Niu} and release the requirements of PLS. The scheme exploits an important fact that, in fountain-coded transmissions, any receiver must obtain sufficient numbers of coded packets to recover the original source information. If the destination can accumulate the packets more quickly than the eavesdropper, the security will be guaranteed. In \cite{Niu}, we noted that the channel fading itself is competent to achieve this goal if the average channel condition of the source-destination link is superior to that of the source-eavesdropper link. Otherwise, transmit power control is executed to yield a higher packet reception rate at the destination. We in this paper attempt to introduce this kind of security concept to resist untrusted relays. Due to the characteristics of fountain coding, however. Eve can still decode a small number of original packets before enough fountain packets are obtained. An revised coding scheme to overcome this drawback is thus proposed first. Then, the application of the revised coding scheme in cooperative networks to resist untrusted relays is analyzed for different cooperation protocols. Specifically, Our contributions are listed as follows,
\begin{enumerate}
\item A new secure coding scheme with linear complexity is designed, which avoid the information leakage caused by the fountain codes before enough packets are received correctly. The proposed coding scheme can be used to relax the strict requirements of PLS, and thus the transmission can be realized according to the ordinary channel capacity instead of the secrecy capacity.
\item The security performance (intercept probability) is derived when applying the proposed scheme in the untrusted relay networks. The intercept probability is proved to be reduced exponentially with the number of the original packets. In addition, it is observed that adopting cooperation achieves a better performance compared to treating the cooperative users as pure eavesdroppers. The DBJ scheme is then found to be optimal and its security performance is acceptable even for a larger number of relays.
\end{enumerate}

The rest of the paper is organized as follows. The system model is introduced in Section II. Section III describes the proposed coding scheme and further analyzes its secrecy and complexity performance. The application of the proposed scheme in untrusted relay systems is considered in Section IV and the intercept probabilities for direct transmission, decode-and-forward (DF) protocol, amplify-and-forward (AF) protocol and DBJ are derived. The numerical results are provided to evaluate the theoretical analysis and the superiority of the proposed scheme in Section V. Finally, Section VI concludes the paper.

\section{System Model}

\begin{figure}
\begin{center}
  \includegraphics[width=80 mm]{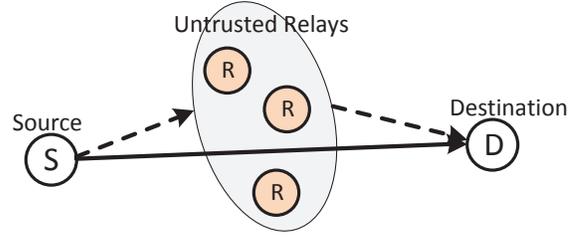}
\end{center}
\caption{Cooperative networks with untrusted relays}
\label{fig:1}       
\end{figure}

The considered cooperative network with untrusted relays is as shown in Fig.\,\ref{fig:1}, which consists of one source (S), one destination (D) and \emph{K} untrusted relays (denoted by a relay set ${\bf{R}}{\rm{ = }}\left\{ {{\rm{R}}_{\rm{1}} {\rm{,R}}_{\rm{2}} {\rm{,}}...{\rm{,R}}_{\rm{K}} } \right\}$). All of the nodes are operated in a half-duplex mode with a single antenna. S intends to deliver a confidential information to D through packet-based wireless transmission with the potential cooperation of the relays. However, the relays are only service level trust but not data level trust \cite{Khodakarami}, i.e., they attempts to intercept (decode) the information simultaneously when they are asked to cooperate. For the worst case, the relays are also supposed to be collusive with each other.

The transmission links are modelled as the block-flat Rayleigh fading channels. That is to say, the channel coefficient $h_{uv}$ between node $u$ and node $v$ is a complex circularly symmetric Gaussian variable with mean zero and variance ${\sigma_{uv} ^2}$. Meanwhile, $h_{uv}$ remains constant during the transmission of one packet and varies independently among different packets. S and the relays (if cooperation is selected) transmit the packets with power $P$ and a target rate $R$ bits per channel use (bpcu). The noise at each receiver is represented by the additive white Gaussian noise with variance $N_0$. Therefore, the received signal-to-noise ratio (SNR) for the link between node $u$ and node $v$ is written as ${\gamma _{uv}} = \rho {\left| {{h_{uv}}} \right|^2}$, where $\rho  = P/{N_0}$. The ${\gamma _{uv}}$ follows an exponential distribution with parameter $\lambda _{uv}=\left( {\rho \sigma _{uv}^2 } \right)^{ - 1}$.

The original packets of the confidential information are denoted by $\left( {I_1 ,I_2 ,...,I_N } \right)$. For the direct transmission, the instantaneous achievable rate at D can be expressed as
\begin{eqnarray}
C_D=\log _2 \left( {1 + \gamma _{SD} } \right)
    \label{eq:exb1}
\end{eqnarray}
Considering that the capacity-approaching code is adopted at the physical layer, D receives $I_n$ correctly if $C_D \ge R$. The system uses basic automatic repeat-request (ARQ) mechanism to ensure D obtain all of the packets, in which the receivers simply discard the erroneous packets. The relays are treated pure eavesdroppers and the achievable rate of the relay ${\rm{R}}_{\rm{k}} $ is
\begin{eqnarray}
C_k=\log _2 \left( {1 + \gamma _{Sk} } \right)
    \label{eq:exb2}
\end{eqnarray}
Since the relays are collusive, they can share their obtained information. For simplicity, we assume that one original packet is intercepted if any relay decodes it correctly and the mutual information accumulation among the relays is not considered. Therefore, the packet $I_n$ is intercepted if $\mathop {\max }\limits_{{\rm{R}}_{\rm{k}} \in {\bf{R}} } \left\{ {C_k } \right\} \ge R$.

If the cooperation mode is decided, the relays assist the transmission of S but also act as the eavesdroppers at the same time. As a low complexity scheme to benefit from the multi-relay cooperation, relay selection strategy is adopted in which only the best relay is selected to cooperate for every transmission. The transmission is divided into two phases. In phase I S broadcasts its packet, and in phase II the selected relay forwards its received information by either DF or AF protocol. The achievable rate at D by using the DF protocol is given by
\begin{eqnarray}
C_D^{DF}=0.5\log _2 \left( {1 + \gamma _{SD}  + \mathop {\max }\limits_{{\rm{R}}_{\rm{k}} \in {\bf{R'}}} \left\{ {\gamma _{kD} } \right\}} \right)
    \label{eq:exb3}
\end{eqnarray}
where, 0.5 is due to the spectral loss of cooperative communications and ${\bf{R'}}$ is the decoding set which includes the relays that succeed in decoding the transmitted packet. As for the AF protocol, the achievable rate at D is
\begin{eqnarray}
C_D^{AF}=0.5\log _2 \left( {1 + \gamma _{SD}  + \mathop {\max }\limits_{{\rm{R}}_{\rm{k}} \in {\bf{R}}} \left\{ {\frac{{\gamma _{Sk} \gamma _{kD} }}{{1 + \gamma _{Sk}  + \gamma _{kD} }}} \right\}} \right)
    \label{eq:exb4}
\end{eqnarray}

The achievable rate at relay ${\rm{R}}_{\rm{k}} $ in the cooperation mode is $C_k' =0.5 C_k$. Therefore, it requires $\mathop {\max }\limits_{{\rm{R}}_{\rm{k}} \in {\bf{R}} } \left\{ {C_k' } \right\} \ge R$ to intercept one packet.

%
%

\section{Information Self-encrypted Scheme for Secure Cooperation with Untrusted Relays}

It can be observed that the relays can easily intercept some confidential information if the original packets are transmitted directly. Therefore, an information self-encrypted scheme is proposed to resist untrusted relays while employing them to cooperate.

\subsection{Scheme description}

Intuitively, the original packets can be pre-processed such that a certain number of the processed packets are necessary to recover the original information. And thus the transmission is secured if the destination receives enough processed packets faster than the eavesdropper. In our previous work \cite{Niu}, the fountain coding scheme is adopted to realized this kind of secure transmission. Due to the characteristics of fountain coding, however, the eavesdropper(s) can still decode a small number of original packets before enough fountain packets are obtained. In this paper, a fixed linear coding scheme is proposed which achieves a much better information self-encrypted performance. Specifically, for the $N$ original packets $\left( {I_1 ,I_2 ,...,I_N } \right)$, the following $N \times N$ generator matrix is designed,
\begin{eqnarray}
T = \left[ \begin{array}{l}
 \begin{array}{*{20}c}
   0 & 1 &
   1 &  \cdots  & 1 & 1  \\
\end{array} \\
 \begin{array}{*{20}c}
   1 & 0 &
   1 &  \cdots  & 1 & 1  \\
\end{array} \\
 \begin{array}{*{20}c}
   1 & 1 &
   0 &  \cdots  & 1 & 1  \\
\end{array} \\
 \begin{array}{*{20}c}
    \,\vdots  &  \;\vdots  &
    \,\vdots  & \, \ddots  & \, \vdots  & \, \vdots   \\
\end{array} \\
 \begin{array}{*{20}c}
   1 & 1 &
   1 &  \cdots  & 0 & 1  \\
\end{array} \\
 \begin{array}{*{20}c}
   1 & 1 &
   1 &  \cdots  & 1 & 0  \\
\end{array} \\
 \end{array} \right]
    \label{eq:exc1}
\end{eqnarray}
Assuming each original packet has $K$ bits, i.e., $I_n  = [i_{1n} ,i_{2n} ,...,i_{Kn} ]^{tr}$, the proposed coding scheme is shown as
\begin{eqnarray}
\left[ {P_1 ,P_2 ,...,P_N } \right] = \left[ {I_1 ,I_2 ,...,I_N } \right] * T(\bmod 2)
    \label{eq:exc2}
\end{eqnarray}
where, the processed packet $P_n  = [p_{1n} ,p_{2n} ,...,p_{Kn}]^{tr}$ also has $K$ bits. The scheme can be represented as the following equation by the bit-wise modulo two operation among the original packets,
\begin{eqnarray}
P_n  = I_1  \oplus ... \oplus I_{n - 1}  \oplus I_{n + 1}  \oplus ... \oplus I_N  = \sum\limits_{l = 1}^N {I_l  \oplus I_n }
    \label{eq:exc3}
\end{eqnarray}

If $N$ is an even number, the generator matrix $T$ has an inverse matrix $T^{-1}=T$. That is to say, the original packets can be recovered through the same process as the encoding scheme,
\begin{eqnarray}
I_n  = P_1  \oplus ... \oplus P_{n - 1}  \oplus P_{n + 1}  \oplus ... \oplus P_N  = \sum\limits_{l = 1}^N {P_l  \oplus P_n }
    \label{eq:exc4}
\end{eqnarray}

\begin{proof}
Let's denote the processed packet set as $\mathbb{P_N}=\{ P_1 ,P_2 ,...,P_N \}$. For a subset with cardinality $M$ ($M  \le N$) $\mathbb{P_M}=\{ P_{m_1} ,P_{m_2} ,...,P_{M} \} \subseteq \mathbb{P_N}$, the complement is $\mathbb{P_M^C}=\mathbb{P_N}-\mathbb{P_M}$. The similar denotations are also applied to the original packet sets $\mathbb{I_N}$, $\mathbb{I_M}$ and $\mathbb{I_M^C}$.

If a
 subset $\mathbb{P_M}$ of the processed packets is received correctly, then
\begin{eqnarray}
\sum {\mathbb{P_M}}&=&P_{m_1}\oplus P_{m_2} \oplus ...\oplus P_{M} \nonumber\\
&=&\left( {M\sum\limits_{l = 1}^N {I_l } } \right) \oplus I_{m_1}  \oplus I_{m_2}  \oplus ... \oplus I_M \nonumber
    \label{eq:exc5}
\end{eqnarray}
where, $\left( {M\sum\limits_{l = 1}^N {I_l } } \right) =\underbrace {\sum\limits_{l = 1}^N {I_l }  \oplus ... \oplus \sum\limits_{l = 1}^N {I_l } }_M$. Because $\left( {2\sum\limits_{l = 1}^N {I_l } } \right) =  {\bf{0}}$, we can obtain $\sum {\mathbb{P_M}}=$
\begin{eqnarray}
\left\{ \begin{array}{l}
 \left( {\sum\limits_{l = 1}^N {I_l } } \right) \oplus I_{m_1}  \oplus I_{m_2}  \oplus ... \oplus I_M  = \sum {\mathbb{I_M^C} } ,M \rm{\; is \; odd}\\
 I_{m_1}  \oplus I_{m_2}  \oplus ... \oplus I_M  = \sum {\mathbb{I_M}} ,M \rm{\; is \; even}\\
 \end{array} \right. \nonumber
    \label{eq:exc6}
\end{eqnarray}

Therefore, if and only if $M$ is an odd number and $N=M+1$, we can derive an individual original packet $\sum {\mathbb{I_M^C} }=\mathbb{I_M^C}=\mathbb{I_N}-\mathbb{I_M}$ since the cardinality of ${\mathbb{I_M^C}} $ is \emph{one} in this case. In other words, we can recover one original packet by $ P_1  \oplus ... \oplus P_{n - 1}  \oplus P_{n + 1}  \oplus ... \oplus P_N =I_n$ if and only if $N$ is an even number.
\end{proof}

If the number of the original packets is odd in practical application, we can simply add a redundant packet to perform the proposed scheme.

\subsection{Secrecy performance and complexity analysis}

\subsubsection{Shannon perfect secrecy}

According to the scheme description, we can get the following conclusion: if less than $N-1$, $N-1$ and $N$ processed packets are obtained correctly, $0$, only $1$ and all of the $N$ original packets can be recovered respectively. As for the recovery of one specified original packet $I_n$, it needs $N-1$ processed packets $\left\{ {P_1,...,P_{n - 1},P_{n + 1},...,P_N} \right\}$. The perfect secrecy of $I_n$ is proved to be achieved if less than these $N-1$ processed packets are received.

\begin{proof}
In \cite{Shannon}, Shannon describes that the perfect secrecy can be achieved if $\Pr [M\left| E \right.] = \Pr [M]$ for all $M$ (set of message) and all $E$ (set of cryptogram). We assume that the source generates bit streams independently with equal probability\footnote{For unideal sources, this can be realized with a scrambler. The information of the scrambler may be publicly known, which does not affect the security performance of the proposed scheme.}, such that the secrecy depends on the security of each bit (bitwise security). The original packets $\left( {I_1 ,I_2 ,...,I_N} \right)$ are consequently also independent of each other. For each bit $i_{kn}$ in the original packet $I_n$, $\Pr [i_{kn}=0]=\Pr [i_{kn}=1]=0.5$. Since $P_n   = \sum\limits_{l = 1}^N {I_l  \oplus I_n }$, we can easily derive that $P_1 ,P_2 ,...,P_N$ are independent as well and for each bit $p_{kn}$ in the processed packet $P_n$, $\Pr [p_{kn}=0]=\Pr [p_{kn}=1]=0.5$.

Denoting $\left\{ P_1,...,P_{n - 1},P_{n + 1},...,P_N \right\}  \buildrel \Delta \over = \mathbb{ P}_{n }^\mathbb{C}$, if a proper subset $\mathbb{ P _J}\subset  \mathbb{ P}_{n }^\mathbb{C}$ is obtained by the receiver, no information about $I_n$ is intercepted. That is because considering any bit $i_{kn}$ in the original packet $I_n$ and the corresponding bit $p_{kJ}$ in the $\sum {\mathbb{ P _J}}$, $\Pr [i_{kn}\left| p_{kJ} \right.] = \Pr [i_{kn}]=0.5$ for all $ i_{kn} \in \left\{ {0,1} \right\}$ and $p_{kJ} \in \left\{ {0,1} \right\}$. That is to say, before all of the packets in $\mathbb{ P}_{n }^\mathbb{C}$ are received correctly, the perfect secrecy of the original packet $I_n$ is maintained. Therefore, all of the original packets are perfectly secured (i.e., the perfect secrecy of the confidential information is ensured) before $N-1$ processed packets are received.
\end{proof}

In case we neglect the information leakage caused by one original packet (only one original packet can be recovered when $N-1$ processed packets are received), it is regarded that the information is intercepted only if the untrusted relays obtain all of the $N$  processed packets when the transmission from S to D is finished. By exploiting the random channel fading at the physical layer, an arbitrarily small intercept probability can be realized as $N$ increases. This characteristic is utilized to resist these untrusted relays and satisfy any predetermined secrecy constraints, which is analyzed in detail in next section.

\subsubsection{Linear complexity}

The encoding process can be designed directly based on Eq. (\ref{eq:exc3}) that $P_n  = \sum\limits_{l = 1}^N {I_l  \oplus I_n }$, in which the $\sum\limits_{l = 1}^N {I_l  }$ is calculated first and then each $P_n $ is derived by the modulo two operation with $I_n$. Therefore, the complexity of the proposed scheme is $2N$ (i.e., linear complexity $\mathcal{O}(N)$). The complexity of the decoding process has the same result.

\section{Intercept Probability of the Proposed Scheme}

In this section, the intercept probability of the proposed secure cooperation scheme is analyzed. The cooperative networks with and without direct link between S and D are both considered.

\subsection{Cooperative networks with direct link}

\subsubsection{Direct transmission}

In this case, the untrusted relays are not employed and simply treated as the eavesdroppers. The probability that one received packet is not decoded correctly at the destination is \cite{Goldsmith}
\begin{eqnarray}
\varepsilon _{D}  &=& \Pr \left\{ {C_D  = {\log _2 \left( {1 + \gamma _{SD} } \right)} < R} \right\} \nonumber\\
&=& 1 - e^{ - \lambda _{SD} \tau}=F\left( {\lambda _{SD} \tau} \right)
    \label{eq:exda1}
\end{eqnarray}
where, $F\left( \chi  \right) = 1 - \exp \left( { - \chi } \right)$ and $\tau=2^R-1$. On the other hand, the relays have the packet error probability
\begin{eqnarray}
\varepsilon _{R}  &=&\Pr \left\{ { \mathop {\max }\limits_{{\rm{R}}_{\rm{k}} \in {\bf{R}} } \left\{ {C_k } \right\} = \log _2 \left( {1 + \mathop {\max }\limits_{{\rm{R}}_{\rm{k}} \in {\bf{R}} } \left\{ {\gamma _{Sk} } \right\}} \right)
 < R} \right\} \nonumber\\
&=&\prod\limits_{k = 1}^K {\left( {1 - e^{ - \lambda _{Sk} \tau} } \right)}=\prod\limits_{k = 1}^K {F\left( {\lambda _{Sk} \tau} \right)}
    \label{eq:exda2}
\end{eqnarray}

The required time slots $T_D$ and $T_R$ for the destination and the relays to recover one packet meets the geometric distribution with parameter $1-\varepsilon _{D}$ and $1-\varepsilon _{R}$ respectively, i.e., their probability mass function (PMF) and cumulative distribution function (CDF) are given by
\begin{eqnarray}
f_v\left( {T_v } \right) = \varepsilon _{v} ^{T_v  - 1}  \left( {1- \varepsilon _{v} } \right)
 \label{eq:ex8}
\end{eqnarray}
\begin{eqnarray}
F_v\left( {T_v } \right) = \sum\limits_{t_v  = 1}^{T_v } { \varepsilon _{v} ^{t_v  - 1}  \left( {1- \varepsilon _{v} } \right)  }
= 1- \varepsilon _{v} ^{T_v}
    \label{eq:exda3}
\end{eqnarray}

To obtain one processed packet, the relays should decode the packet correctly not after it is correctly received by the destination through ARQ. Thus, the intercept probability for one processed packet is calculated as
\begin{eqnarray}
 &&P_{I - 1}= \sum\limits_{T_D  = 1}^\infty  {f_D \left( {T_D } \right)F_R \left( {T_D } \right)}  \nonumber\\
 && = \sum\limits_{T_D  = 1}^\infty  {\varepsilon _D^{T_D  - 1} (1 - \varepsilon _D )(1 - \varepsilon _R^{T_D } )} = \frac{{1 - \varepsilon _R }}{{1 - \varepsilon _D \varepsilon _R }}
    \label{eq:exda4}
\end{eqnarray}

In order to intercept the confidential information, the eavesdropper should receives all of the $\left( {P_1 ,P_2 ,...,P_N} \right)$ correctly. The intercept probability is thus derived as
\begin{eqnarray}
P_I  = \left( {P_{I - 1} } \right)^N
    \label{eq:exda5}
\end{eqnarray}
It is observed that the intercept probability decreases to zero \emph{exponentially} as \emph{N} increases, which means that any arbitrary small intercept probability can be satisfied by simply increasing the value of \emph{N}.

\subsubsection{DF protocol}

For the DF cooperation, there should be some processed packets that D receives directly without the cooperation of any relays (i.e., the decoding set ${\rm{R'}}$ should be null set). In this case, the packet error probability of D is
\begin{eqnarray}
&&\hspace{-7mm}\varepsilon _{D}^{DF}\nonumber\\
&&\hspace{-7mm}= \Pr \left\{ {C_D^{DF}= 0.5\log _2 \left( {1 + \gamma _{SD}  + \mathop {\max }\limits_{{\rm{R}}_{\rm{k}} \in \emptyset } \left\{ {\gamma _{kD} } \right\}} \right) < R} \right\} \nonumber\\
&&\hspace{-7mm}= 1 - e^{ - \lambda _{SD} \tau ' } =F\left( {\lambda _{SD} \tau '} \right)
    \label{eq:exdb1}
\end{eqnarray}
where, $\tau '=2^{2R}-1$. Simultaneously, $\varepsilon _{R}^{DF}=\prod\limits_{k = 1}^K {F\left( {\lambda _{Sk} \tau'} \right)}$. The secure transmission of one packet is thus obtained as
\begin{eqnarray}
 P_{S - 1}^{DF} & =& \sum\limits_{T_D  = 1}^\infty  {f_D \left( {T_D } \right)\left( {1 - F_R \left( {T_D } \right)} \right)}  \nonumber\\
  &=& \sum\limits_{T_D  = 1}^\infty  {\left( {\varepsilon _D^{DF} } \right)^{T_D  - 1} (1 - \varepsilon _D^{DF} )\left( {\varepsilon _R^{DF} } \right)^{T_D } }  \nonumber\\
  &=& \frac{{{\varepsilon _R^{DF} }  - \varepsilon _D^{DF} {\varepsilon _R^{DF} } }}{{1 - \varepsilon _D^{DF} {\varepsilon _R^{DF} } }}
    \label{eq:exdb2}
\end{eqnarray}
Therefore, the intercept probability of one packet is
\begin{eqnarray}
P_{I - 1}^{DF}  = 1 - P_{S - 1}^{DF}  = \frac{{1 - {\varepsilon _R^{DF} } }}{{1 - \varepsilon _D^{DF} {\varepsilon _R^{DF} } }}
    \label{eq:exdb3}
\end{eqnarray}

The intercept probability of the confidential information is also
\begin{eqnarray}
P_I^{DF}  = \left( {P_{I - 1}^{DF} } \right)^N
    \label{eq:exdb4}
\end{eqnarray}

For mathematical convenience and fair comparison among different numbers of relays, it is assumed that for all of the $\lambda _{Sk}$ and $\lambda _{kD}$ are identical respectively in this paper, and denoted that $\lambda _{Sk}=\lambda _{SR}$ and $\lambda _{kD}=\lambda _{RD}$. We can easily derive
\begin{eqnarray}
P_{I - 1}^{DF} (K_1 ) > P_{I - 1}^{DF} (K_2 ),\quad {\rm{for}} \  K_1  > K_2
    \label{eq:exdb5}
\end{eqnarray}
from the fact that $\varepsilon _{R}(K_1 ) < \varepsilon _{R}(K_2 )$. Therefore, exploiting more untrusted relays \emph{deteriorates} the secrecy performance, although more diversity gain can be achieved. There is a tradeoff consideration in practical applications between the secrecy and diversity performance when employing untrusted relays. Furthermore, whether to select the cooperation or not can be also decided by comparing Eq. (\ref{eq:exda4}) and Eq. (\ref{eq:exdb3}). However, we can still reduce the intercept probability \emph{exponentially} by increasing the value of \emph{N}.

\subsubsection{AF protocol}

It is intractable to obtain the exact intercept probability of the AF protocol. Therefore, we consider its upper and lower bounds by using the following inequalities,
\begin{eqnarray}
 \frac{{\gamma _{Sk} \gamma _{kD} }}{{1 + \gamma _{Sk}  + \gamma _{kD} }} \ge \frac{{\gamma _{Sk} \gamma _{kD} }}{{\gamma _{Sk}  + \gamma _{kD} }}{\rm{ - }}\frac{1}{4} \nonumber\\
  \ge \frac{1}{2}\min \left\{ {\gamma _{Sk} ,\gamma _{kD} } \right\}{\rm{ - }}\frac{1}{4} \buildrel \Delta \over = \gamma _{AF}^ -
    \label{eq:exdc1}
\end{eqnarray}
and,
\begin{eqnarray}
\frac{{\gamma _{Sk} \gamma _{kD} }}{{1 + \gamma _{Sk}  + \gamma _{kD} }} \le \frac{{\gamma _{Sk} \gamma _{kD} }}{{\gamma _{Sk}  + \gamma _{kD} }} \le \min \left\{ {\gamma _{Sk} ,\gamma _{kD} } \right\}  \buildrel \Delta \over = \gamma _{AF}^ +
    \label{eq:exdc2}
\end{eqnarray}
where the first equality comes from the result derived in \cite{Behnad}.

First the upper bound of the intercept probability is analyzed. We can derive the probability that both D and the untrusted relays cannot receive a packet correctly as follows,
\begin{eqnarray}
&&\hspace{-10mm} \varepsilon _{R,D}^{AF+} =\Pr \left\{ {\mathop {\max }\limits_{k \in {\rm{R}}} \left\{ {\gamma _{Sk} } \right\} < R,\gamma _{SD}  + \mathop {\max }\limits_{k \in {\rm{R}}} \left\{ {\gamma _{AF}^ -  } \right\} < R} \right\} \nonumber\\
 &&\hspace{-10mm}=G\left( {F\left( {\lambda _{SR} \tau' } \right),\frac{{\lambda _{SR} }}{{I_1 }}F\left( {I_1 \tau' } \right)-A_1 } \right. \nonumber\\
&&\hspace{-7mm} + \left. { \frac{{\lambda _{RD} }}{{I_1- \lambda _{SR}}} \left[ { F\left( {\lambda _{SR} \tau' } \right)- \frac{{\lambda _{SR} }}{{   I_1 }}F\left( { I_1\tau' } \right)} \right] -A_2 } \right) 
    \label{eq:exdc3}
\end{eqnarray}

where,
\begin{eqnarray}
 &&\hspace{-5mm}G\left( {\alpha ,\beta } \right) = K\sum\limits_{k_1  = 0}^{K - 1} {\sum\limits_{k_2  = 0}^{K - 1 - k_1 } {\left[ {\left( \begin{array}{l}
 K - 1 \\
\hspace{3mm} k_1  \\
 \end{array} \right)} \right.} }  \nonumber\\
&&\hspace{-5mm} \left. {\left( \begin{array}{l}
 K - 1 - k_1  \\
\hspace{7mm} k_2  \\
 \end{array} \right)\alpha ^{k_1 } \left( { - 1} \right)^{k_2 } \left( {1 - \alpha } \right)^{K - 1 - k_1  - k_2 } \beta } \right] \nonumber
    \label{eq:exdc4}
\end{eqnarray}
$I_1  = \left( {\lambda _{SR}  + \lambda _{RD} } \right)\left( {k_2  + 1} \right) + \lambda _{RD} \left( {K - 1 - k_1  - k_2 } \right)$, $I_2  = I_1  - 0.5\lambda _{SD}$, $\tau_1  = 2^{2R}  - 0.75$ and,
\begin{eqnarray}
A_1 = \left\{ \begin{array}{l}
 {\lambda _{SR} \left[ {1 - F(\lambda _{SD} \tau_1 )} \right]}\tau', I_2=0 \\
 \frac{{\lambda _{SR} \left[ {1 - F(\lambda _{SD} \tau_1 )} \right]}}{{I_2 }}F\left( {I_2 \tau' } \right), else \\
 \end{array} \right. \nonumber\\
 A_2 = \left\{ \begin{array}{l}
\frac{{\lambda _{RD} \left[ {1 - F(\lambda _{SD} \tau_1 )} \right]}}{{\lambda _{SR} }}  \left[ { F(\lambda _{SR}\tau' )} \right.\\
 \hspace{5mm} \left. { -\lambda _{SR}\tau'\exp \left( { - \lambda _{SR}\tau' } \right) } \right], I_2-\lambda _{SR}=0 \\
 \frac{{\lambda _{RD} \left[ {1 - F(\lambda _{SD} \tau_1 )} \right]}}{{I_2-\lambda _{SR} }}\left[ {F\left( {\lambda _{SR} \tau' } \right) -\lambda _{SR}\tau'} \right] \\
 \hspace{40mm}, I_2=0 \\
 \frac{{\lambda _{RD} \left[ {1 - F(\lambda _{SD} \tau_1 )} \right]}}{{I_2-\lambda _{SR} }}\left[ {F\left( {\lambda _{SR} \tau' } \right) } \right.\\
 \hspace{30mm} \left. { -\frac{{\lambda _{SR} }}{{   I_2 }}F\left( {   I_2 \tau' } \right)} \right],else \\
 \end{array} \right. \nonumber
\end{eqnarray}
(See Appendix A).

Then the probability that only D decodes a packet correctly is similarly calculated as,
\begin{eqnarray}
&&\hspace{-2mm}\varepsilon _{R,\rlap{--} D}^{AF+}=\Pr \left\{ {\mathop {\max }\limits_{k \in {\rm{R}}} \left\{ {\gamma _{Sk} } \right\} < R,\gamma _{SD}  + \mathop {\max }\limits_{k \in {\rm{R}}} \left\{ {\gamma _{AF}^ -  } \right\} \ge R} \right\} \nonumber\\
&&\hspace{-2mm} = G\left( {F\left( {\lambda _{SR} \tau' } \right), A_1 + A_2} \right) 
    \label{eq:exdc5}
\end{eqnarray}

Therefore, the intercept probability of one processed packet can be derived as
\begin{eqnarray}
P_{I - 1}^{AF+}  = 1 - \sum\limits_{T_D  = 1}^\infty  {\left( {\varepsilon _{R,D}^{AF+} } \right)^{T_D  - 1} } \varepsilon _{R,\rlap{--} D}^{AF+}  = 1 - \frac{{\varepsilon _{R,\rlap{--} D}^{AF+} }}{{1 - \varepsilon _{R,D}^{AF+} }}
    \label{eq:exdc6}
\end{eqnarray}
And we can obtain the upper bound of the intercept probability of the confidential information by
\begin{eqnarray}
P_I^{AF+}  = \left( {P_{I - 1}^{AF+} } \right)^N
    \label{eq:exdc7}
\end{eqnarray}
On the other hand, the expression of the lower band $P_{I}^{AF-}$ is similar to $P_{I}^{AF+}$, and we only need to replace $\tau_1$ with $\tau'$ and set $I_2=I_1-\lambda _{SD}$. Therefore, the intercept probability of the confidential information is also reduced \emph{exponentially} as the value of \emph{N} increases. The upper bound of the intercept probability can be used to decide a suitable value of \emph{N} to satisfy any required security level. It is not intuitive to compare the intercept probabilities for different numbers of relays, and we will observe the tendency through the numerical results.

\subsection{Cooperative networks without direct link}

In some practical networks, the direct link between S and D is blocked such that the transmission is realized only through the help of relays. However, it is impossible to realize the secure transmission by DF protocol since the relays should decode the packets correctly to forward them. For the ordinary AF protocol, it is also unusable because for any relay ${\rm{R}}_{\rm{k}} $ we have $\gamma _{Sk}  \ge \frac{{\gamma _{Sk} \gamma _{kD} }}{{1 + \gamma _{Sk}  + \gamma _{kD} }}$. The relays can always decode the packets correctly if D does. Therefore, a scheme called destination-based jamming (DBJ) is widely adopted in literatures. In this paper, we also combines the DBJ with our proposed scheme to realize the secure transmission. In DBJ, the transmit power $P$ in Phase I is allocated between S and D with parameter $\alpha  \in [1,0]$. S transmits its packet with power $\alpha P$ and D transmits artificial noise with power $\left( {1 - \alpha } \right)P$ simultaneously. Then the relay forwards the received super-position signal still with power $P$. D can subtract the artificial noise from the received signal since it is generated by itself, while the relays are confused by it. Therefore, the achievable rate at the relays and D is derived respectively as \cite{Sun1},
\begin{eqnarray}
C_k^{DBJ}  = 0.5\log _2 \left( {1 + \frac{{\alpha \gamma _{Sk} }}{{1 + \left( {1 - \alpha } \right)\gamma _{kD} }}} \right)
    \label{eq:exdd1}
\end{eqnarray}
\begin{eqnarray}
&&\hspace{-8mm} C_D^{DBJ}  = 0.5\log _2 \left( {1 + \mathop {\max }\limits_{{\rm{R}}_{\rm{k}} \in {\bf{R}}} \left\{ {\frac{{\alpha \gamma _{Sk} \gamma _{kD} }}{{1 + \alpha \gamma _{Sk}  + \left( {2 - \alpha } \right)\gamma _{kD} }}} \right\}} \right)\nonumber\\
&&\hspace{-8mm}
    \label{eq:exdd2}
\end{eqnarray}

First, we calculate the upper bound of the intercept probability based on the following equations

\begin{eqnarray}
&& \frac{{\alpha \gamma _{Sk} \gamma _{kD} }}{{1 + \alpha \gamma _{Sk}  + \left( {2 - \alpha } \right)\gamma _{kD} }} \nonumber\\
&&\hspace{-3mm} \ge \frac{{1}}{{\left( {2 - \alpha } \right)}}\left( {\frac{1}{2}\min \left\{ {\alpha \gamma _{Sk} ,\left( {2 - \alpha } \right)\gamma _{kD} } \right\} - \frac{1}{4}} \right)
    \label{eq:exdd3}
\end{eqnarray}
\begin{eqnarray}
\frac{{\alpha \gamma _{Sk} }}{{1 + \left( {1 - \alpha } \right)\gamma _{kD} }} < \frac{{\alpha \gamma _{Sk} }}{{\left( {1 - \alpha } \right)\gamma _{kD} }}
    \label{eq:exdd4}
\end{eqnarray}

Let's denote $X_1  = \alpha \gamma _{Sk}$ and $X_2  = \left( {2 - \alpha } \right)\gamma _{kD}$. If $\tau_2  = \left( {2^{2R}  - 1} \right)\frac{{1 - \alpha }}{{2 - \alpha }} \ge 1$, we can derive that
\begin{eqnarray}
&&\hspace{-18mm} \varepsilon _{R,D}^{DBJ + }  = \Pr \left\{ {\mathop {\max }\limits_{{\rm{R}}_{\rm{k}} \in {\bf{R}}} \left\{ {\frac{{X_1 }}{{X_2 }}} \right\} < \tau_2 ,} \right. \nonumber\\
&&\hspace{0mm} \left. {\mathop {\max }\limits_{{\rm{R}}_{\rm{k}} \in {\bf{R}}} \left\{ {\min \left\{ {X_1 ,X_2 } \right\}} \right\} < \tau_3 } \right\} \nonumber\\
&&\hspace{-7mm}  = G\left( {\frac{{\lambda _{X_2 } }}{{\lambda _{X_2 }  + \lambda _{X_1 } \tau_2 }},\frac{{\lambda _{X_1 } }}{{I_3 }}F\left( {I_3 \tau_3 } \right)} \right. \nonumber\\
&&\hspace{0mm} \left. { + \frac{{\lambda _{X_2 } }}{{I_3  }}F\left( {I_3 \tau_3  } \right) - \frac{{\lambda _{X_2 } }}{{I_{4} }}F\left( {I_{4} \tau_3 } \right)} \right)
    \label{eq:exdd5}
\end{eqnarray}
and
\begin{eqnarray}
&&\hspace{-15mm}  \varepsilon _{R,\rlap{--} D}^{DBJ + }  = \Pr \left\{ {\mathop {\max }\limits_{{\rm{R}}_{\rm{k}} \in {\bf{R}}} \left\{ {\frac{{X_1 }}{{X_2 }}} \right\} < \tau_2 ,} \right. \nonumber\\
&& \left. {\mathop {\max }\limits_{{\rm{R}}_{\rm{k}} \in {\bf{R}}} \left\{ {\min \left\{ {X_1 ,X_2 } \right\}} \right\} \ge \tau_3 } \right\} \nonumber\\
&&\hspace{-4mm}   = G\left( {\frac{{\lambda _{X_2 } }}{{\lambda _{X_2 }  + \lambda _{X_1 } \tau_2 }},\frac{{\lambda _{X_1 } }}{{I_3 }}\left[ {1 - F\left( {I_3 \tau_3  } \right)} \right]} \right. \nonumber\\
&& \left. { + \frac{{\lambda _{X_2 } }}{{I_3 }}\left[ {1 - F\left( {I_3 \tau_3 } \right)} \right] - \frac{{\lambda _{X_2 } }}{{I_{4} }}\left[ {1 - F\left( {I_{4} \tau_3 } \right)} \right]} \right)
    \label{eq:exdd6}
\end{eqnarray}
where $\tau_3   = 2\left[ {\left( {2^{2R}  - 1} \right)\left( {2 - \alpha } \right) + 0.25} \right]$, $\lambda _{X_1 }  = \lambda _{SR} \alpha ^{ - 1}$, $\lambda _{X_2 }  = \lambda _{RD} \left( {2 - \alpha } \right)^{ - 1}$, $I_3  = \left( {\lambda _{X_1 }  + \lambda _{X_2 } } \right)\left( {k_2  + 1} \right) + \left( {\lambda _{X_2 }  + \lambda _{X_1 } \tau_2 } \right)\left( {K - 1 - k_1  - k_2 } \right)$ and $I_{4}  = \lambda _{X_1 } \tau_2  + \lambda _{X_2 }  + \left( {\lambda _{X_1 }  + \lambda _{X_2 } } \right)k_2  + \left( {\lambda _{X_2 }  + \lambda _{X_1 } \tau_2 } \right)\left( {K - 1 - k_1  - k_2 } \right)$.

On the other hand, if $\tau_2 < 1$,
\begin{eqnarray}
&&\hspace{-7mm} \varepsilon _{R,D}^{DBJ + }   = G'\left( {1,1,\left( {\frac{{\lambda _{X_1 } }}{{\lambda _{X_1 }  + \lambda _{X_2 } /\tau_2 }}} \right)^{K - 1} \frac{{\lambda _{X_1 } }}{{I_{5} }}\left[ {F\left( {I_{5} \tau_3 } \right)} \right.} \right)\nonumber\\
    \label{eq:exdd7}
\end{eqnarray}
and
\begin{eqnarray}
&&\hspace{-8mm}\varepsilon _{R,\rlap{--} D}^{DBJ + }\nonumber\\
&&\hspace{-8mm}  = G'\left( {1,1,\left( {\frac{{\lambda _{X_1 } }}{{\lambda _{X_1 }  + \lambda _{X_2 } /\tau_2 }}} \right)^{K - 1} \frac{{\lambda _{X_1 } }}{{I_{5} }}\left[ {1 - F\left( {I_{5} \tau_3 } \right)} \right]} \right)\nonumber\\
    \label{eq:exdd8}
\end{eqnarray}
where,

\noindent $G'\left( {\varsigma ,\beta ,\varphi } \right) = K\sum\limits_{k_1  = 0}^{K - 1} {\left[ {\left( \begin{array}{l}
 K - 1 \\
 k_1  \\
 \end{array} \right)\left( { - \varsigma } \right)^{k_1 } \beta ^{K - 1 - k_2 } \varphi } \right]}$, $I_5=\left( {{\lambda _{{X_1}}} + {\lambda _{{X_2}}}/\tau_2} \right)\left( {{k_1} + 1} \right)$,

Therefore, the upper bound of the intercept probability of the confidential information is
\begin{eqnarray}
P_I  = \left( {P_{I - 1}^{DBJ+} } \right)^N  = \left( {1 - \frac{{\varepsilon _{R,\rlap{--} D}^{DBJ+} }}{{1 - \varepsilon _{R,D}^{DBJ+} }}} \right)^N
    \label{eq:exdd9}
\end{eqnarray}

To derive the lower bound of the intercept probability, the following inequality is utilized
\begin{eqnarray}
 &&\hspace{5mm} \frac{{\alpha \gamma _{Sk} \gamma _{kD} }}{{1 + \alpha \gamma _{Sk}  + \left( {2 - \alpha } \right)\gamma _{kD} }}  \nonumber\\
  &&\le \frac{1}{{\left( {2 - \alpha } \right)}}\min \left\{ {\alpha \gamma _{Sk} ,\left( {2 - \alpha } \right)\gamma _{kD} } \right\} \nonumber\\
 &&  \le \left\{ \begin{array}{l}
  \gamma _{kD} ,
\textcircled{\small{1}}\; {\rm{if}} \; \alpha \sigma _{Sk}^2  \ge \left( {2 - \alpha } \right)\sigma _{kD}^2 \\
 \frac{{\alpha \gamma _{Sk} }}{{\left( {2 - \alpha } \right)}},
\textcircled{\small{2}}\; {\rm{if}} \; \alpha \sigma _{Sk}^2  < \left( {2 - \alpha } \right)\sigma _{kD}^2
 \end{array} \right.
    \label{eq:exdd10}
\end{eqnarray}

For case $\textcircled{\small{1}}$, we can derived that
\begin{eqnarray}
 &&\hspace{-5mm} \varepsilon _{R,D}^{DBJ- }=  \nonumber\\
  &&\hspace{-5mm}\Pr \left\{ {\mathop {\max }\limits_{{\rm{R}}_{\rm{k}} \in {\bf{R}}} \left\{ {\frac{{\alpha \gamma _{Sk} }}{{1 + \left( {1 - \alpha } \right)\gamma _{kD} }}} \right\} < \tau' ,\mathop {\max }\limits_{{\rm{R}}_{\rm{k}} \in {\bf{R}}} \left\{ {\gamma _{kD} } \right\} < \tau' } \right\} \nonumber\\
  &&\hspace{-5mm} = G\left( {1 - \frac{{\lambda _{RD} e^{ - \lambda _{SR} \tau' /\alpha } }}{{I_6 }},\frac{{\lambda _{RD} }}{{I_7 }}F\left( {I_7 \tau' } \right) } \right. \nonumber\\
  &&\hspace{-2mm} \left. {-\frac{{\lambda _{RD} e^{ - \lambda _{SR} \tau' /\alpha } }}{{I_8 }}F\left( {I_8 \tau' } \right)} \right)
    \label{eq:exdd11}
\end{eqnarray}
and,
\begin{eqnarray}
&&\hspace{-5mm} \varepsilon _{R,\rlap{--} D}^{DBJ - }  =\nonumber\\
&&\hspace{-5mm} \Pr \left\{ {\mathop {\max }\limits_{k \in {\rm{R}}} \left\{ {\frac{{\alpha \gamma _{Sk} }}{{1 + \left( {1 - \alpha } \right)\gamma _{kD} }}} \right\} < \tau' ,\mathop {\max }\limits_{k \in {\rm{R}}} \left\{ {\gamma _{kD} } \right\} \ge \tau' } \right\} \nonumber\\
&&\hspace{-5mm}  = G\left( {1 - \frac{{\lambda _{RD} e^{ - \lambda _{SR} \tau' /\alpha } }}{{I_6 }},\frac{{\lambda _{RD} }}{{I_7 }}\left[ {1 - F\left( {I_7 \tau' } \right)} \right] } \right. \nonumber\\
&&\hspace{-2mm} \left. {-\frac{{\lambda _{RD} e^{ - \lambda _{SR} \tau' /\alpha } }}{{I_8 }}\left[ {1 - F\left( {I_8 \tau' } \right)} \right]} \right)
    \label{eq:exdd12}
\end{eqnarray}
where, $I_6  = \lambda _{RD}  + \lambda _{SR} \left( {1 - \alpha } \right)\tau'/\alpha$, $I_7  = \lambda _{RD} \left( {k_2  + 1} \right) + I_6 \left( {K - 1 - k_1  - k_2 } \right)$ and $I_8  = I_7  + \lambda _{SR} \left( {1 - \alpha } \right)\tau'/\alpha$.

Now the attentions turn to the case $\textcircled{\small{2}}$, we can derive that
\begin{eqnarray}
&&\hspace{-12mm} \varepsilon _{R,D}^{DBJ - }  = \Pr \left\{ {\mathop {\max }\limits_{k \in {\rm{R}}} \left\{ {\frac{{\alpha \gamma _{Sk} }}{{1 + \left( {1 - \alpha } \right)\gamma _{kD} }}} \right\} < \tau' ,} \right. \nonumber\\
&&\hspace{3mm} \left. {\mathop {\max }\limits_{k \in {\rm{R}}} \left\{ {\frac{{\alpha \gamma _{Sk} }}{{\left( {2 - \alpha } \right)}}} \right\} < \tau' } \right\} \nonumber\\
&&\hspace{-12mm}  = G'\left( {1,1,\frac{1}{{\left( {k_1  + 1} \right)}}F\left( {\frac{\left( {k_1  + 1} \right)\lambda _{SR}{\tau' }}{{\alpha }}} \right)} \right)+ \nonumber\\
&&\hspace{-12mm}  G'\left( {I_{10} ,I_{11} ,\frac{{\lambda _{SR} e^{{\frac{{\lambda _{SD} }}{{\left( {1 - \alpha } \right)}}}} }}{{I_9 \left( {k_1  + 1} \right)}}\left[ {F\left( {\frac{{I_9 \left( {k_1  + 1} \right)\left( {2 - \alpha } \right)\tau'}}{\alpha }} \right)} \right.} \right. \nonumber\\
&&\hspace{-12mm} \left. {\left. { - F\left( {\frac{{I_9 \left( {k_1  + 1} \right)\tau' }}{\alpha }} \right)} \right]} \right)
    \label{eq:exdd13}
\end{eqnarray}
and
\begin{eqnarray}
&&\hspace{-15mm} \varepsilon _{R,\rlap{--} D}^{DBJ - }  = \Pr \left\{ {\mathop {\max }\limits_{k \in {\rm{R}}} \left\{ {\frac{{\alpha \gamma _{Sk} }}{{1 + \left( {1 - \alpha } \right)\gamma _{kD} }}} \right\} < \tau' ,} \right. \nonumber\\
&& \left. {\mathop {\max }\limits_{k \in {\rm{R}}} \left\{ {\frac{{\alpha \gamma _{Sk} }}{{\left( {2 - \alpha } \right)}}} \right\} \ge \tau'} \right\} \nonumber\\
&&\hspace{-3mm}  = G'\left( {I_{10} ,I_{11} ,\frac{{\lambda _{SR} e^{\frac{{\lambda _{SD} }}{{\left( {1 - \alpha } \right)}}} }}{{I_9 \left( {k_1  + 1} \right)}}  } \right. \nonumber\\
&&\left. { \left[ {1-F\left( {\frac{{I_9 \left( {k_1  + 1} \right)\left( {2 - \alpha } \right)\tau'}}{\alpha }} \right)} \right] } \right)
    \label{eq:exdd14}
\end{eqnarray}
where,
$I_9  = \lambda _{SR}  + \frac{{\lambda _{RD} \alpha }}{{\left( {1 - \alpha } \right)\tau' }}$, $I_{10}  = \frac{{\lambda _{SR} e^{\frac{{\lambda _{RD} }}{{\left( {1 - \alpha } \right)}}} }}{{I_9 }}$ and $I_{11}  = F\left( {\frac{{\lambda _{SR} \tau' }}{\alpha }} \right) + I_{10} \left( {1- F\left( {\frac{{I_9 \tau' }}{\alpha }} \right)} \right)$.

The lower bound of the intercept probability for both cases $\textcircled{\small{1}}$ and $\textcircled{\small{2}}$ is
\begin{eqnarray}
P_I^{DBJ - }  = \left( {P_{I - 1}^{DBJ - } } \right)^N  = \left( {1 - \frac{{\varepsilon _{R,\rlap{--} D}^{DBJ - } }}{{1 - \varepsilon _{R,D}^{DBJ - } }}} \right)^N
    \label{eq:exdd15}
\end{eqnarray}

For the cooperative networks without direct link, any required security level can be also satisfied by increasing the value of \emph{N}.


\section{Numerical Results and Discussions}
In this section, the numerical results are provided to validate the theoretical analysis first, and then some remarkable conclusions are discussed. The simulation environment is established in a rectangular coordinate system, where the source and the destination are located at (0,0) and (0,1) respectively. The position of the relays is generated between the source and the destination. Without loss of generality, $\rho$ is set to be 20dB and $R=1$bit/s/Hz.

\begin{figure}
\begin{center}
  \includegraphics[width=90 mm]{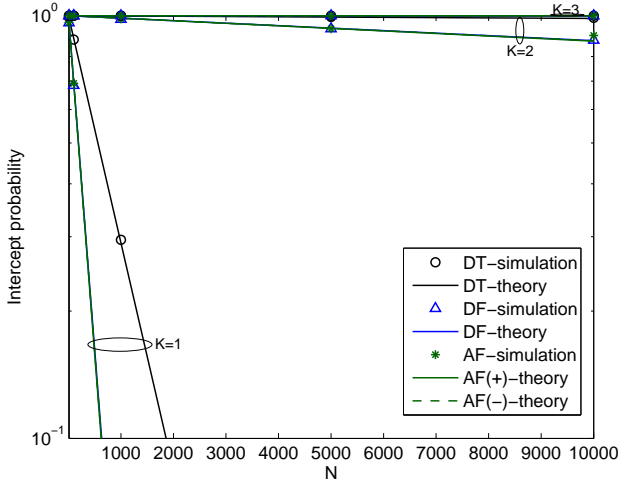}
\end{center}
\caption{Intercept probability vs. \emph{N} for the cooperative networks with direct link}
\label{fig:2}       
\end{figure}

The cooperative network with direct link is considered in Fig.\,\ref{fig:2} for $K=1, 2 and 3$, where the locations of the relays is assumed to be the midpoint between the source and the destination, i.e., (0.5,0). It can be observed that exploiting the relays to assist the transmission achieves a better security performance compared to treating them as the pure eavesdroppers. More importantly, the intercept probability reduces to zero exponentially with the values of \emph{N} by using our proposed scheme, which can be used to realize the required security level and is the main superiority compared to the alternative schemes. However, in accordance with the existing literatures \cite{Sun1,Khodakarami,Sun2}, the security performance is deteriorated significantly with more relays (eavesdroppers). For $K \ge 2$ it is almost impossible to make a secure transmission due to the diversity gain at the relays, which makes us expect the results of DBJ.

\begin{figure}
\begin{center}
  \includegraphics[width=90 mm]{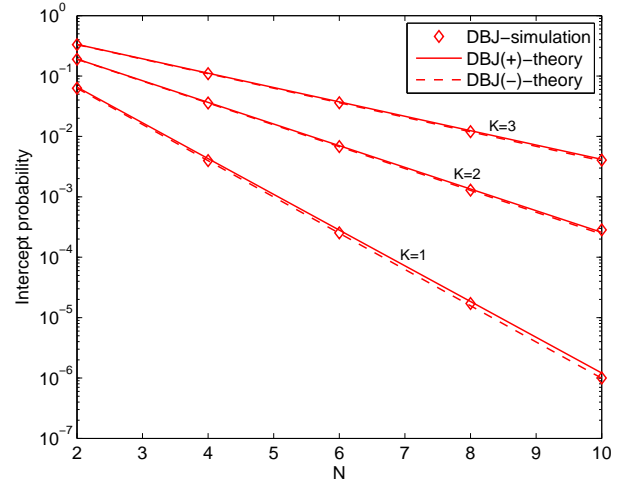}
\end{center}
\caption{Intercept probability vs. \emph{N} for the cooperative networks without direct link}
\label{fig:3}       
\end{figure}

\begin{figure}
\begin{center}
  \includegraphics[width=90 mm]{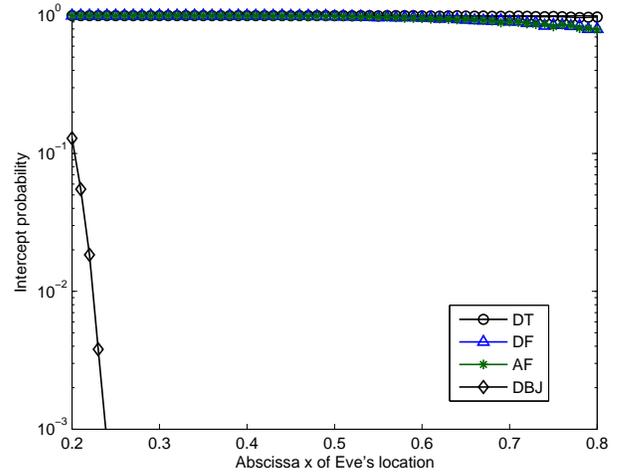}
\end{center}
\caption{Intercept probability vs. \emph{N} for different locations of relays ($K=2$ and $N=1000$)}
\label{fig:4}       
\end{figure}

Fig.\,\ref{fig:3} illustrates the results of intercept probability for the cooperative networks without direct links, in which the DBJ strategy is adopted \footnote{Because of the simulation time, only the results with small \emph{N} values are given. For the results with large \emph{N} values, the intercept probability can be obtained based on the exponential decline principle.}. Although the security performance is still deteriorated with the number of the relays (eavesdroppers), adopting DBJ results in interference-limited effect on the relays and thus improves the security performance dramatically. Even for a large number of relays (eavesdroppers), the intercept probability is acceptable by combining DBJ with our proposed scheme for large \emph{N} values. The intercept probability with different locations of relays ($K=2$ and $N=1000$) is given in Fig.\,\ref{fig:4}. It is observed that the intercept probability is increased as the relays approaches to the source, which is also an inherent weaknesses of the physical layer security. However, DBJ still outperforms much better than the conventional cooperation scheme and achieves an acceptable security performance.

\begin{figure}
\begin{center}
  \includegraphics[width=90 mm]{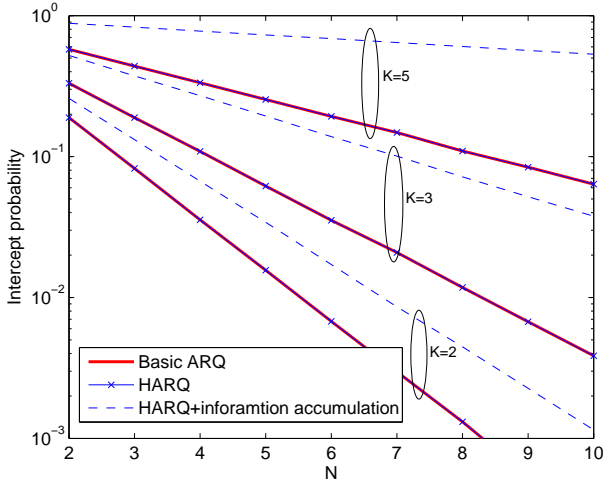}
\end{center}
\caption{Intercept probability vs. N for the cooperative networks with HARQ receivers}
\label{fig:5}       
\end{figure}

Due to its significant advantage, only the DBJ is considered in the following discussions. First, the simulation results considering HARQ at the receivers is given in Fig.\,\ref{fig:5}. The intercept probability is nearly the same between the Basic ARQ and HARQ receivers. The one reason is that both the relays and the destination can benefit from the HARQ protocol. In addition, the worst case that the information accumulation among the relays is also considered. Although the information accumulation among the relays deteriorate the security performance, the intercept probability is still reduced to zero as \emph{N} increases.

\begin{figure}
\begin{center}
  \includegraphics[width=90 mm]{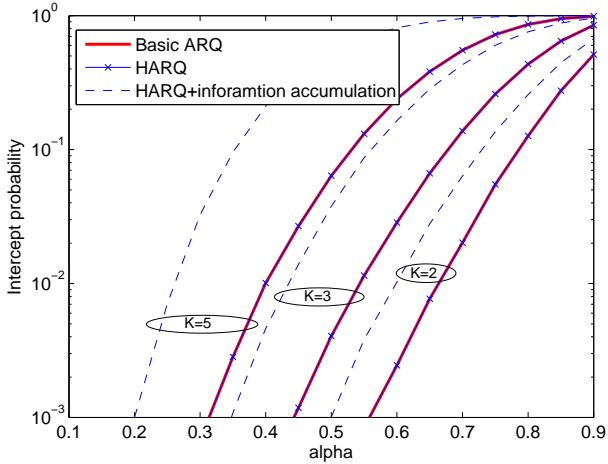}
\end{center}
\caption{Intercept probability vs. N for the cooperative networks with different alpha values}
\label{fig:6}       
\end{figure}

In Fig.\,\ref{fig:6}, the effect of the alpha on the intercept probability is shown. It is observed that the intercept probability is decreased by reducing the value of alpha (i.e., increasing the interference caused by the jamming signal at the relays). However, reducing the value of alpha means more power is provided to generate the jamming signal and thus less power is available for the information transmission, such that the information transmission needs more time slots. To satisfy both the transmission efficiency and security requirement, the values of alpha and \emph{N} should be designed jointly. 

\section{Conclusion}

An information self-encrypted scheme is proposed in this paper to resist untrusted relays in the cooperative networks. The idea of the proposed scheme is to utilized the original packets to encrypt each other as the secret key, such that the receivers need to receive all of the processed packets to recover the original information. The security is realized if the eavesdropper cannot obtain all of the processed packets. The application of this scheme in the untrusted relay networks is then analyzed. It is shown that the intercept probability is reduced exponentially with the number of the original packets, and user cooperation achieves a better performance than direct transmission. However, the conventional cooperation does not work well for a large number of relays. To solve this problem, the DBJ strategy is also combined with our proposed scheme and a more acceptable security performance is realized. The numerical results are then given to confirm the analysis. The proposed scheme can be also combined with other physical layer techniques, e.g., adaptive power allocation and beamforming, to further improve the security performance. It may be also a valuable topic to consider the proposed scheme with PLS jointly.


%

\appendices
\section{Derivation of $\varepsilon _{R,D}^{AF+}$}

Denoting the relay selected to cooperate is ${\dot k}$, the derivation of $\varepsilon _{R,D}^{AF}$ can be divided into two parts: 1) $\gamma _{S\dot k}  \le \gamma _{\dot kD}$; 2) $\gamma _{S\dot k}  > \gamma _{\dot kD}$, and
\begin{eqnarray}
\varepsilon _{R,D}^{AF+}  = \varepsilon _{R,D(1)}^{AF+}  + \varepsilon _{R,D(2)}^{AF}
    \label{eq:exA1}
\end{eqnarray}
Correspondingly,
\begin{eqnarray}
&&\hspace{-7mm} \varepsilon _{R,D(1)}^{AF+} = \nonumber\\
&&\hspace{-7mm}  \sum\limits_{\dot k = 1}^K {\Pr \left\{ {\mathop {\max }\limits_{k \in {\rm{R}}} \left\{ {\gamma _{Sk} } \right\} < \tau' ,\gamma _{SD}  + \frac{1}{2}\gamma _{S\dot k}  < \tau_1 ,\gamma _{S\dot k}  \le \gamma _{\dot kD} } \right\}} \nonumber\\
&&\hspace{-7mm}  = \sum\limits_{\dot k = 1}^K {\int_0^{\tau' } {\lambda _{S\dot k} e^{ - \lambda _{S\dot k} x} e^{ - \lambda _{\dot kD} x} \left[ {1 - e^{ - \lambda _{SD} (\tau_1  - \frac{1}{2}x)} } \right]} }  \nonumber\\
&&\hspace{-3mm} \prod\limits_{k' \in \left\{ {1,...,K} \right\} - \left\{ {\dot k} \right\}} {\left( {\int_0^x {\lambda _{Sk'} e^{ - \lambda _{Sk'} y} e^{ - \lambda _{k'D} y} dy + } } \right.}  \nonumber\\
&&\hspace{-3mm} \left. {\int_0^x {\lambda _{k'D} e^{ - \lambda _{k'D} y} \int_y^{\tau' } {\lambda _{Sk'} e^{ - \lambda _{Sk'} z} } dzdy} } \right)dx \nonumber\\
&&\hspace{-7mm}  = K\int_0^{\tau' } {\lambda _{SR} e^{ - \lambda _{SR} x} e^{ - \lambda _{RD} x} \left[ {1 - e^{ - \lambda _{SD} (\tau_1  - \frac{1}{2}x)} } \right]}  \nonumber\\
&&\hspace{-7mm} \left[ {1 - e^{ - \lambda _{SR} \tau' }  - e^{ - \left( {\lambda _{SR}  + \lambda _{RD} } \right)x}  + e^{ - \lambda _{SR} \tau'  - \lambda _{RD} x} } \right]^{K - 1} dx \nonumber\\
 &&\hspace{-7mm}=G\left( {F\left( {\lambda _{SR} \tau' } \right),\frac{{\lambda _{SR} }}{{I_1 }}F\left( {I_1 \tau' } \right) } - A_1 \right) \nonumber\\
    \label{eq:exA2}
\end{eqnarray}
where,
\begin{eqnarray}
 &&\hspace{-5mm}G\left( {\alpha ,\beta } \right) = K\sum\limits_{k_1  = 0}^{K - 1} {\sum\limits_{k_2  = 0}^{K - 1 - k_1 } {\left[ {\left( \begin{array}{l}
 K - 1 \\
\hspace{3mm} k_1  \\
 \end{array} \right)} \right.} }  \nonumber\\
&&\hspace{-5mm} \left. {\left( \begin{array}{l}
 K - 1 - k_1  \\
\hspace{7mm} k_2  \\
 \end{array} \right)\alpha ^{k_1 } \left( { - 1} \right)^{k_2 } \left( {1 - \alpha } \right)^{K - 1 - k_1  - k_2 } \beta } \right] \nonumber
    \label{eq:exA3}
\end{eqnarray}
$F\left( \chi  \right) = 1 - \exp \left( { - \chi } \right)$, $I_1  = \left( {\lambda _{SR}  + \lambda _{RD} } \right)\left( {k_2  + 1} \right) + \lambda _{RD} \left( {K - 1 - k_1  - k_2 } \right)$, $I_2  = I_1  - 0.5\lambda _{SD}$, $ \tau_1 = 2^{2R}  - 0.75$ and,
\begin{eqnarray}
A_1 = \left\{ \begin{array}{l}
 {\lambda _{SR} \left[ {1 - F(\lambda _{SD} \tau_1 )} \right]}\tau', I_2=0 \\
 \frac{{\lambda _{SR} \left[ {1 - F(\lambda _{SD} \tau_1 )} \right]}}{{I_2 }}F\left( {I_2 \tau' } \right), else \\
 \end{array} \right. \nonumber
\end{eqnarray}

Similarly, we can derive that
\begin{eqnarray}
&&\hspace{-7mm} \varepsilon _{R,D(2)}^{AF+} = \nonumber\\
&&\hspace{-7mm}  \sum\limits_{\dot k = 1}^K {\Pr \left\{ {\mathop {\max }\limits_{k \in {\rm{R}}} \left\{ {\gamma _{Sk} } \right\} < \tau' ,\gamma _{SD}  + \frac{1}{2}\gamma _{\dot kD}  < \tau_1 ,\gamma _{S\dot k}  > \gamma _{\dot kD} } \right\}} \nonumber\\
&&\hspace{-7mm}= G\left( {F\left( {\lambda _{SR} \tau' } \right),\frac{{\lambda _{RD} }}{{I_1- \lambda _{SR}}} \left[ { F\left( {\lambda _{SR} \tau' } \right)} \right. } \right. \nonumber\\
&&\hspace{-3mm}\left. { \left. { - \frac{{\lambda _{SR} }}{{   I_1 }}F\left( { I_1\tau' } \right)} \right] -A_2} \right)  \nonumber\\
    \label{eq:exA4}
\end{eqnarray}
where,
\begin{eqnarray}
A_2 = \left\{ \begin{array}{l}
\frac{{\lambda _{RD} \left[ {1 - F(\lambda _{SD} \tau_1 )} \right]}}{{\lambda _{SR} }}  \left[ { F(\lambda _{SR}\tau' )} \right.\\
 \hspace{5mm} \left. { -\lambda _{SR}\tau'\exp \left( { - \lambda _{SR}\tau' } \right) } \right], I_2-\lambda _{SR}=0 \\
 \frac{{\lambda _{RD} \left[ {1 - F(\lambda _{SD} \tau_1 )} \right]}}{{I_2-\lambda _{SR} }}\left[ {F\left( {\lambda _{SR} \tau' } \right) -\lambda _{SR}\tau'} \right] \\
 \hspace{40mm}, I_2=0 \\
 \frac{{\lambda _{RD} \left[ {1 - F(\lambda _{SD} \tau_1 )} \right]}}{{I_2-\lambda _{SR} }}\left[ {F\left( {\lambda _{SR} \tau' } \right) } \right.\\
 \hspace{30mm} \left. { -\frac{{\lambda _{SR} }}{{   I_2 }}F\left( {   I_2 \tau' } \right)} \right],else \\
 \end{array} \right. \nonumber
\end{eqnarray}

Then, the $\varepsilon _{R,D}^{AF}$ is obtained by combining Eq. (\ref{eq:exA1}), Eq. (\ref{eq:exA2}) and Eq. (\ref{eq:exA4}).

%
%
%
%

\ifCLASSOPTIONcaptionsoff
  \newpage
\fi


\begin{thebibliography}{1}

\bibitem{Sendonaris}
Sendonaris, A., Erkip, E., and Aazhang, B. (2003). User cooperation diversity-part I: system description. \emph{IEEE Transactions on Communications}, 51(11), 1927-1938.

\bibitem{Laneman}
Laneman, J. N., Tse, D. N. C., and Wornell, G. W. (2004). Cooperative diversity in wireless networks: efficient protocols and outage behavior. \emph{IEEE Transactions on Information Theory}, 50(12), 3062-3080.

\bibitem{Wyner}
Wyner, A. D. (1975). The wire-tap channel. \emph{Bell System Technical Journal}, 54(8), 1355-1387.

\bibitem{Csiszar}
Csiszar, I., and Korner, J. (1978). Broadcast channels with confidential messages. \emph{IEEE Transactions on Information Theory},  24(3), 339-348.

\bibitem{Barros} Barros, J., and Rodrigues, M. R. D. (2006). Secrecy capacity of wireless channels. in \emph{Proceedings of IEEE ISIT '06}, pp. 356--360.

\bibitem{Gopala}
Gopala, P. K., Lai, L., and Gamal, H. EI. (2008). On the secrecy capacity of fading channels. \emph{IEEE Transactions on Information Theory}, 54(10), 4687 - 4698.

\bibitem{Saad}
Saad, W., Zhou, X., Han, Z., and Poor, H. V. (2014). On the physical layer security of backscatter wireless systems. \emph{IEEE Transactions on Wireless Communications}, 13(6), 3442-3451.

\bibitem{Yang}
Yang, N., Wang, L., Geraci, G., Elkashlan, M., Yuan, J., and Renzo, M. D. (2015). Safeguarding 5G wireless communication networks using physical layer security. \emph{IEEE Communications Magazine}, 53(4), 20-27.

\bibitem{Krikidis1}
Krikidis, I., Thompson, J. S., and McLaughlin, S. (2009). Relay selection for secure cooperative networks with jamming. \emph{IEEE Transactions on Wireless Communications}, 8(10), 5003-5011.

\bibitem{Dong}
Dong, L., Han, Z., Petropulu, A., and Poor, H. V. (2010). Improving wireless physical layer security via cooperating relays. \emph{IEEE Transactions on Signal Processing}, 58(3), 1875-1888.

\bibitem{Bassily}
Bassily, R., and Ulukus, S. (2013). Deaf cooperation and relay selection strategies for secure communication in multiple relay networks. \emph{IEEE Transactions on Signal Processing}, 61(6), 1544-1554.

\bibitem{Hui} Hui, H., Swindlehurst, A. L., Li, G., and Liang, J. (2015).
		Secure relay and jammer selection for physical layer security. \emph{IEEE Signal Processing Letters}, 22(8). 1147--1151.

\bibitem{Krikidis2}
Krikidis, I. (2010). Opportunistic relay selection for cooperative networks with secrecy constraints. \emph{IET Communications}, 4(15), 1787-1791.

\bibitem{Zou1} Zou, Y., Wang, X., and Shen, W. (2013).
		Optimal relay selection for physical-layer security in cooperative wireless networks. \emph{IEEE Journal on Selected Areas in Communications}, 31(10), 2099-2111.

\bibitem{Fan}
Fan, L., Lei,X., Duong, T. Q.,Elkashlan, M., and Karagiannidis, G. K. (2014). Secure multiuser communications in multiple amplify-and-forward relay networks. \emph{IEEE Transactions on Communications}, 62(9), 3299-3310.

\bibitem{Zou2}
Zou, Y., Zhu, J., Wang, X., and Leung, V. C. M. (2015). Improving physical-layer security in wireless communications using diversity techniques. \emph{IEEE Network}, 29(1), 42-48.

\bibitem{He1}
X. He, and A. Yener, ��Two-hop secure communication using an untrusted relay: A case for cooperative jamming,�� in Proc. IEEE Globecom, New Orleans, LA, Dec. 2008, pp. 1�C5.

\bibitem{He2}
X. He, and A. Yener, ��Cooperation with an untrusted relay: A secrecy perspective,�� IEEE Trans. Inf. Theory, vol. , no. 8, pp. 3807�C3827, Aug. 2010.

\bibitem{Jeong}
C. Jeong, I.-M. Kim, and D. I. Kim, ��Joint secure beamforming design at the source and the relay for an amplify-and-forward MIMO untrusted relay system,�� IEEE Trans. Signal Process., vol. 60, no. 1, pp. 310�C325, Jan. 2012.

\bibitem{Huang}
J. Huang, A. Mukherjee, and A. L. Swindlehurst, ��Secure communication via an untrusted non-regenrative relay in fading channels,�� IEEE Trans. Signal Process., vol. 61, no. 10, pp. 2536�C2550, May 2013.

\bibitem{Wang}
L. Wang, M. Elkashlan, J. Huang, N. H. Tran, and T. Q. Duong, ��Secure transmission with optimal power allocation in untrusted relay networks,�� IEEE Wireless Commun. Lett., vol. 3, no. 3, pp. 289�C292, Jun. 2014.

\bibitem{Sun1}
L. Sun, T. Zhang, Y. Li, and H. Niu, ��Performance study of two-hop amplify-and-forward systems with untrustworthy relay nodes,�� IEEE Trans. Veh. Technol., vol. 61, no. 8, pp. 3801�C3807, Oct. 2012.

\bibitem{Khodakarami}
H. Khodakarami, and F. Lahouti, ��Link adaptation with untrusted relay assignment: Design and performance analysis,�� IEEE Trans. Commun., vol. 61, no. 12, pp. 4874�C4883, Dec. 2013.

\bibitem{Sun2}
L. Sun, P. Ren, Q. Du, Y. Wang, and Z. Gao, ��Security-Aware Relaying Scheme for Cooperative Networks
With Untrusted Relay Nodes,�� IEEE Commun. Lett., vol. 19, no. 3, pp. 463�C466, Mar. 2015.

\bibitem{Niu}
H. Niu, M. Iwai, K. Sezaki, L. Sun, and Q. Du, "Exploiting fountain codes for secure wireless delivery," IEEE Commun. Lett., vol. 18, no. 5, pp. 777-780, May 2014.

\bibitem{Shannon}
C. E. Shannon, "Communication Theory of Secrecy Systems," Bell Systems Technical Journal,. Vol. 28, pp. 656�C715, 1948.

\bibitem{Goldsmith}
A. Goldsmith, Wireless Communications. Cambridge University Press, 2005.

\bibitem{Behnad}
A. Behnad, R. Parseh, and H. Khodakarami, ��Upper bound for the performance metrics of amplify-and-forward cooperative networks based on harmonic mean approximation,�� in Proc. 18th ICT, Ayia Napa, Cyprus, May 2011, pp. 157�C161.

\end{thebibliography}
\end{document}